\documentclass[reprint,amsmath,amssymb,prl,aps,superscriptaddress]{revtex4-1}
\usepackage{txfonts}
\usepackage{pifont}
\usepackage{amssymb}
\usepackage{dcolumn}
\usepackage{amsmath}
\usepackage[dvips]{epsfig}
\usepackage{array}
\usepackage{color}
\usepackage{ulem}

\makeatletter
\def\btt#1{\texttt{\@backslashchar#1}}%
\DeclareRobustCommand\bblash{\btt{\@backslashchar}}%
\makeatother

\begin{document}
\bibliographystyle{apsrev4-1}
\title{Highly reproducible superconductivity in potassium-doped triphenylbismuth}
\author{Ren-Shu Wang}
\affiliation{School of Materials Science and Engineering, Faculty of Physics and Electronic Technology,
Hubei University, Wuhan 430062, China}
\affiliation{Center for High Pressure Science and Technology Advanced Research, Shanghai 201203, China}
\author{Jia Cheng}
\affiliation{School of Materials Science and Engineering, Faculty of Physics and Electronic Technology,
Hubei University, Wuhan 430062, China}
\author{Xiao-Lin Wu}
\affiliation{School of Materials Science and Engineering, Faculty of Physics and Electronic Technology,
Hubei University, Wuhan 430062, China}
\author{Hui Yang}
\affiliation{School of Materials Science and Engineering, Faculty of Physics and Electronic Technology,
Hubei University, Wuhan 430062, China}
\author{Xiao-Jia Chen}
\email{xjchen@hpstar.ac.cn}
\affiliation{Center for High Pressure Science and Technology Advanced Research, Shanghai 201203, China}
\author{Yun Gao}
\email{gaoyun@hubu.edu.cn}
\affiliation{School of Materials Science and Engineering, Faculty of Physics and Electronic Technology,
Hubei University, Wuhan 430062, China}
\author{Zhong-Bing Huang}
\email{huangzb@hubu.edu.cn}
\affiliation{School of Materials Science and Engineering, Faculty of Physics and Electronic Technology,
Hubei University, Wuhan 430062, China}
\date{\today}

\begin{abstract}
Using a new two-step synthesis method - ultrasound treatment and low temperature annealing, we explore superconductivity
in potassium-doped triphenylbismuth, which is composed of one bismuth atom and three phenyl rings. The combination of dc
and ac magnetic measurements reveals that one hundred percent of synthesized samples exhibit superconductivity at 3.5~K
and/or 7.2~K at ambient pressure. The magnetization hysteresis loops provide a strong evidence of type-II superconductor,
with the upper critical magnetic field up to 1.0~Tesla. Both calculated electronic structure and measured Raman spectra
indicate that superconductivity is realized by transferring electron from potassium to carbon atom. Our study opens an
encouraging window for the search of organic superconductors in organometallic molecules.

PACS number(s):{74.70.Kn, 74.25.-q, 78.30.Jw}
\end{abstract}

\maketitle

\textit{Introduction--}
Superconductivity in organic materials has been attracting great attention due to its fundamental importance and potential
application prospect. Following the discovery of superconductivity in (TMTSF)$_2$PF$_6$ in 1980~\cite{Jerome1980}, several
organic superconducting (SC) families have been reported including charge transfer complexes~\cite{Ishiguro1997},
fullerides~\cite{Hebard1991}, and graphites~\cite{Emery2005,Kim2007}. In 2010, potassium doped
picene was shown to display a SC transition temperature T$_c$ as high as 18~K~\cite{Mitsuhashi2010}, which provides a platform
to explore superconductivity in organic hydrocarbons. Soon thereafter, potassium doped phenanthrene~\cite{XFWang2011} and
dibenzopentacene~\cite{Xue2012} were found to exhibit superconductivity at 5~K and 33~K, respectively.
The above three molecules belong to fused hydrocarbons, in which five, three, and seven phenyl rings are fused via sharing
sides. Very recently, we found that by doping potassium into $p$-terphenyl, a hydrocarbon formed by connecting three phenyl
rings with C-C bond, SC transitions can be observed at 123~K, 43~K, and 7.2~K~\cite{RSWang2017}. The observation of distinct
T$_c$ in these hydrocarbons suggests that both the number and connecting pattern of phenyl rings play important roles in the
SC property.

Despite tremendous efforts by the scientific community~\cite{Kambe2012,Ruff2013,Heguri2014,Artioli2015,Heguri2015,YGao2016},
the detailed crystal structures of hydrocarbon superconductors have not yet been determined in experiments so far, due to
low reproducibility of SC samples and vanishingly small SC fraction. This places a serious restriction on deep understanding
of their physical properties. To make progress on hydrocarbon superconductors, we develop a two-step synthesis method -
ultrasound treatment and low temperature annealing to explore superconductivity in potassium-doped triphenylbismuth.
As a member of organometallic molecules~\cite{Elschenbroich2006}, triphenylbismuth has been used as the solidifying catalyst
for butyl hydroxyl propellant of high combustion velocity, as well as the catalyst for some monomers'
polymerization~\cite{Dotterl2012}. In each triphenylbismuth molecule, three phenyl rings and one bismuth atom are connected
by single C-Bi bond. Such an arrangement of phenyl rings is distinct from the ones in fullerides~\cite{Hebard1991},
graphites~\cite{Emery2005,Kim2007}, fused hydrocarbons~\cite{Mitsuhashi2010,XFWang2011,Xue2012},
and $p$-terphenyl~\cite{RSWang2017}. In view of these facts, exploration of superconductivity in potassium-doped
triphenylbismuth not only enriches the functionalities of organometallic compounds but also provides a new platform for
understanding the relationship between SC property and molecular structure.

Potassium-doped triphenylbismuth was synthesized by the following steps. High-purity potassium metal (99\% purity,
Sinopharm Chemical Reagent) was cut into small pieces and mixed with triphenylbismuth ($>$98\% purity, Tokyo Chemical Industry)
with a mole ratio of $x:1$ ($x$=1, 2, 2.5, 3 and 3.5). The mixtures were then loaded into quartz tubes and sealed under high
vacuum ($1\times 10^{-4}$ Pa). The sample tubes were treated in an ultrasound device at 90~$^{o}$C for 10~hours. After
ultrasound treatment, the samples tubes were heated at 130~$^{o}$C for 1-5~days. Here, ultrasound treatment was adopted to mix
potassium and triphenylbismuth thoroughly, and low temperature annealing
can avoid producing KH via reaction of potassium and hydrogen,
which is crucial for crystallization of doped materials. For each run of experiment, the sample from the same tube was distributed
into several nonmagnetic capsules and sealed by GE varnish in a glove box with the oxygen and moisture levels less than 0.1~ppm.

\begin{figure*}
\includegraphics[width=0.95\textwidth]{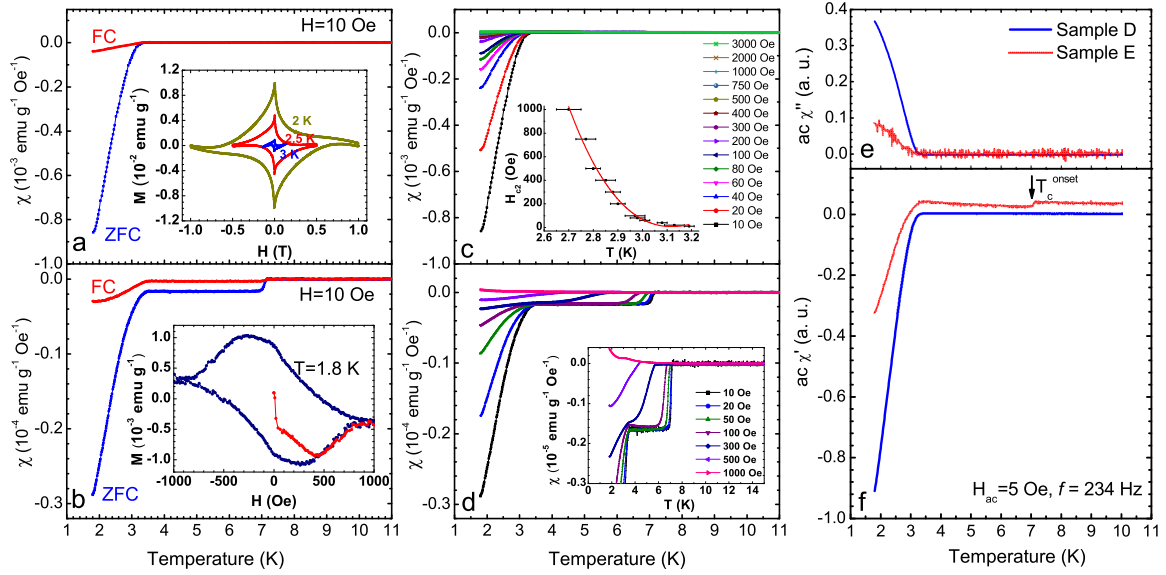}
\caption{(Color online) \textbf{a, b}, The temperature dependence of the dc magnetic susceptibility $\chi$ for samples D and E in
the applied magnetic field of 10~Oe with field cooling (FC) and zero-field cooling (ZFC). The inset figures show the magnetization
loops of samples D and E, with measured temperature shown beside the corresponding loop. \textbf{c, d}, The temperature dependence
of $\chi$ for samples D and E measured at various magnetic fields in the ZFC run. The inset figure in \textbf{c} shows the upper
critical magnetic field H$_{c2}$ in the temperature region of 2.7-3.2~K. The inset figure in \textbf{d} displays the transition
around 7.2~K with enlarged scale. \textbf{e, f}, Imaginary $\chi^{\prime\prime}$ and real $\chi^{\prime}$ components of the ac magnetic
susceptibility as a function of temperature. The probe harmonic magnetic field and frequency are 5~Oe and 234~Hz, respectively.}
\label{Figure1}
\end{figure*}

\textit{Magnetic property of potassium-doped triphenylbismuth--}
Superconductivity was revealed by both the dc and ac magnetic measurements performed on our samples with a SQUID magnetometer
(Quantum Design MPMS3). Pristine triphenylbismuth exhibits a weak diamagnetic behavior, which is clearly characterized by the
small negative magnetic susceptibility in the temperature range of 1.8-300~K. Upon doping potassium into triphenylbismuth,
all synthesized samples listed in Table~\ref{Table1} exhibit a SC transition temperature of $\sim$3.5~K, and four samples also
show a transition at 7.2~K. The representative results for samples D ($x$=3, annealed for 3 days) and E ($x$=2, annealed for 5 days)
are summarized in Fig.~\ref{Figure1}. Figure~\ref{Figure1}a shows the dc magnetic susceptibility $\chi$ for sample D
in the applied magnetic field of 10~Oe with field cooling (FC) and zero-field cooling (ZFC) in the temperature range of 1.8-11~K.
Both FC and ZFC susceptibilities show a sudden decrease around 3.5~K. Such a sudden drop of $\chi$ is consistent with the well-defined
Meissner effect, supporting the occurrence of superconductivity in sample D. The shielding volume fraction at 1.8~K is estimated to
be about 1.7\%, and larger values are expected with further lowering the temperature since $\chi$ does not saturate at 1.8~K.
Similarly, sudden drops of $\chi$ around 3.5 and 7.2~K indicate that there exist two SC phases in sample E (Fig.~\ref{Figure1}b).
Notice that the drop at 7.2~K is much weaker than the one at 3.5~K, implying that sample E is dominated by the SC
phase with T$_c$$\sim$3.5~K.

\begin{table}
\caption{\label{Table1} List of K$_x$triphenylbismuth samples synthesized in this study. Both T$_{c}^{onset}$ and T$_c$ were
read out from the ZFC run measured in the applied magnetic field of 10~Oe. The former denotes the temperature where the magnetic
susceptibility turns to suddenly decrease with lowering the temperature, and the latter is determined from the intercept of
linear extrapolations from below and above T$_{c}^{onset}$.}

\begin{tabular}{c c c c c}
\hline
\hline
Sample label &$x$ &Annealing time (days) & T$_{c}^{onset}$ & T$_c$ \\
\hline
A  & 3 & 1 & 3.39       & ~~3.19       \\
B  & 1 & 3 & 3.06       & ~~2.90       \\
C  & 2 & 3 & 3.56~\&~7.28 & ~~3.35~\&~7.19 \\
\textbf{D}&\textbf{3} &\textbf{3} &\textbf{3.49} & ~~\textbf{3.32}  \\
\textbf{E}&\textbf{2} &\textbf{5} &\textbf{3.51~\&~7.18} & ~~\textbf{3.29~\&~7.13}\\
F  &2.5& 5 & 3.46       & ~~3.30       \\
G  & 3 & 5 & 3.52~\&~7.17 & ~~3.32~\&~7.06 \\
H  &3.5& 5 & 3.53~\&~7.25 & ~~3.28~\&~7.13 \\
\hline
\hline
\end{tabular}
\end{table}

The inset of Fig.~\ref{Figure1}a shows the magnetization loops of sample D with magnetic field up to 1.0~T measured at 2, 2.5, and
3~K in the SC state. The hysteresis loops along the two opposite magnetic field directions show a clear diamond-like shape,
providing a strong evidence for the type-II superconductor. One can readily see that the dip or peak of the magnetization loops
appears at magnetic field close to 0~T, indicating a very small lower critical magnetic field H$_{c1}$ for the SC phase
at 3.5~K. The expansion of the diamond from 3 K to 2~K reflects the fact that the upper critical magnetic field H$_{c2}$ increases
with lowering the temperature. The type-II SC behavior is also applied for sample E, as seen from the magnetization
loop with magnetic field up to 1000~Oe measured at 1.8~K in the inset of Fig.~\ref{Figure1}b. One significant difference from the
magnetization loops of sample D is that the diamond shape is strongly distorted, due to the coexistence of two SC phases in sample E.

The obtained superconductivity in potassium-doped triphenylbismuth was also supported by the evolution of the $\chi$-T curve with
the applied magnetic fields (Fig.~\ref{Figure1}c, Fig.~\ref{Figure1}d and the inset of Fig.~\ref{Figure1}d). The $\chi$-T curve
gradually shifts towards the lower temperatures with increasing magnetic field. This character is consistent with the intrinsic
property of a superconductor, i.e., the SC transition temperature T$_c$ is gradually decreased with increasing magnetic field.
In the inset of Fig.~\ref{Figure1}c, we show the temperature dependence of the upper critical magnetic field H$_{c2}$ for sample D
in the temperature region of 2.7-3.2~K. Here, H$_{c2}$ is determined from the $\chi$-T curves measured at various magnetic fields.
A dramatic increase of H$_{c2}$ with lowering the temperature is clearly evidenced in the investigated temperature region.

The ac magnetic susceptibility measurements were adopted to make a further confirmation for the observed superconductivity. This
technique has been successfully used to study numerous superconductors including high-T$_c$ cuprates~\cite{Gomory1997},
heavy-fermion material CeCu$_2$Si$_2$~\cite{Lengyel2011}, and iron-based FeSe$_{1-x}$~\cite{Bendele2010}. The real component
$\chi^\prime$ of the ac susceptibility is a measure of the magnetic shielding and the imaginary component $\chi^{\prime\prime}$
reflects the magnetic irreversibility~\cite{Gomory1997}. Figures~\ref{Figure1}e and \ref{Figure1}f present the temperature
dependence of $\chi^{\prime\prime}$ and $\chi^\prime$ of the ZFC ac susceptibility, respectively. For sample D (blue lines),
two inflection anomalies occur in the $\chi^{\prime\prime}$-T and $\chi^\prime$-T curves upon cooling at the exactly same temperature
of 3.5~K, which coincides with the T$_c$ value already determined from Fig.~\ref{Figure1}a. As can be seen, both $\chi^\prime$ and
$\chi^{\prime\prime}$ are close to zero above the transition due to the
absence of flux exclusion in the normal state. Upon entering the SC state below 3.5~K, the diamagnetic behavior leads
to a negative $\chi^\prime$ which becomes more negative as more flux is expelled from the sample with lowering the temperature.
Here, a finite $\chi^{\prime\prime}$ reflects the fact that the flux penetrating the sample lags the external flux.
Similar inflection anomalies are also observed for sample E (red lines) around 3.5~K. However, only $\chi^\prime$ exhibits
an anomaly for the SC transition at 7.2~K and no visible change can be found for $\chi^{\prime\prime}$. This is attributed to
the small fraction of the 7.2~K SC phase.

\begin{figure}
\includegraphics[width=0.48\textwidth]{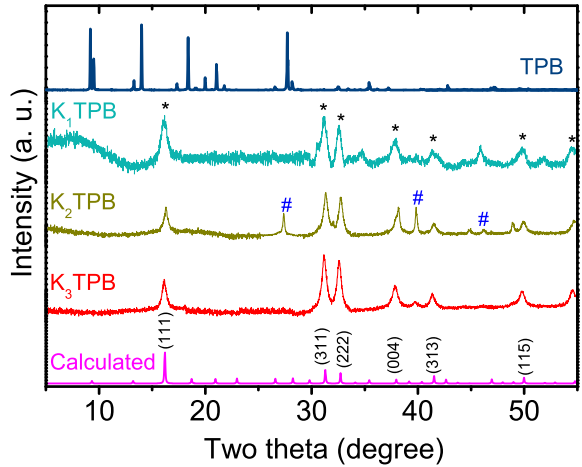}
\caption{(Color online) XRD patterns of pristine and potassium-doped triphenylbismuth measured at room temperature. In figures~\ref{Figure2}
and \ref{Figure5}, TPB, K$_1$TPB, K$_2$TPB, and K$_3$TPB correspond to pristine triphenylbismuth, samples B, E, and D, respectively.
The symbol $*$ represents the common XRD pattern for the three samples, and the symbol \# stands for the solitary XRD pattern for sample E.
The purple curve at the bottom represents the calculated XRD pattern of optimized structure in Fig.~\ref{Figure3}b.}
\label{Figure2}
\end{figure}

\begin{figure}
\includegraphics[width=0.48\textwidth]{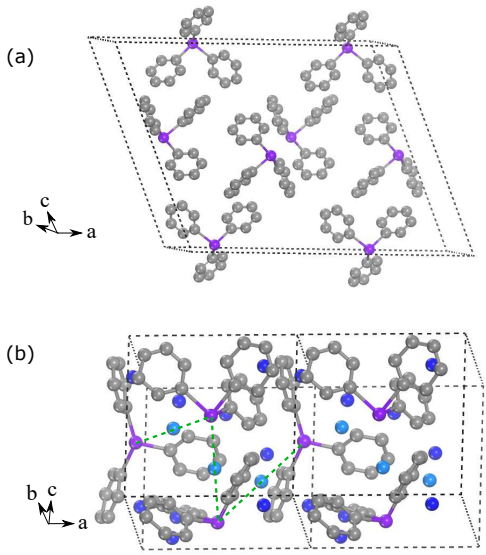}
\caption{(Color online) \textbf{(a)} The molecular arrangement in pristine triphenylbismuth is shown in a single cell; \textbf{(b)} The arrangement
of molecules and potassium in doped material is shown in a $2\times 1\times 1$ supercell. The grey, purple, and blue balls represent carbon, bismuth,
and potassium atoms, respectively. The hydrogen atoms are not given in the figure for clarity.}
\label{Figure3}
\end{figure}

\textit{Crystal structures of pristine and potassium-doped triphenylbismuth--}
X-ray diffraction (XRD) spectrometer (Panalytical Emperean) was employed to examine the evolution of the crystal structure from pristine
to potassium-doped triphenylbismuth. Figure~\ref{Figure2} displays the XRD patterns of pristine and potassium-doped samples with mole ratio
$x$=1, 2, and 3. The peak positions for the pristine material are in good agreement with the ones in the standard PDF card. Pristine
triphenylbismuth is a typical kind of molecular solid and crystallizes in the space group C2/c (No.~15), with eight molecules of
C$_{18}$H$_{15}$Bi in a unit cell of dimensions a=27.70~$\AA$, b=5.82~$\AA$, c=20.45~$\AA$, and $\beta$=114.48$^o$~\cite{Hawley1968,Jones1995},
as shown in Fig.~\ref{Figure3}a. The mean Bi-C distance is 2.24~$\AA$ and the mean C-Bi-C bond angle is about 94$^o$. Upon doping potassium,
no obvious peak appears at the positions where the pristine material shows strong peaks and the XRD feature is completely different from
the undoped case. This indicates that doping of potassium atoms produces a new crystal structure.

Notice that the samples with different $x's$ exhibit a common XRD pattern marked by $*$, while the sample with $x$=2 also shows
a solitary pattern marked by \#, which is actually consistent with the XRD pattern of metal Bi. The existence of Bi in sample E is evidenced
by the appearance of colorless and transparent liquid on the inner wall of tube, indicating that partial triphenylbismuth molecules are
decomposed into Bi atoms and phenyls. Given that all the three samples exhibit superconductivity around 3.5~K, it is reasonable to ascribe
the 3.5~K SC phase to the crystal structure represented by $*$. As to the 7.2~K SC phase observed in sample E,
its very small fraction makes it hard to discern the corresponding crystallization information from the XRD results.

\begin{figure}
\includegraphics[width=0.48\textwidth]{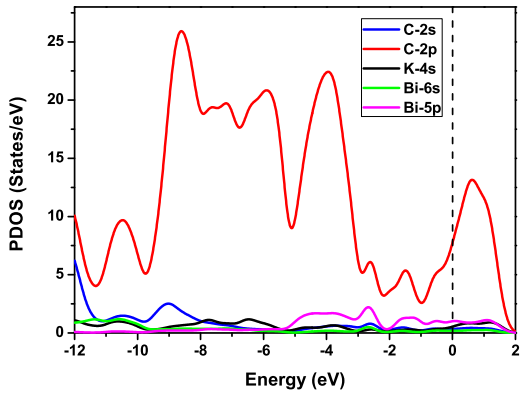}
\caption{(Color online) Orbital-resolved partial density of states (PDOS) as a function of energy. The Fermi energy is set to be zero.
The red, blue, black, green, and purple solid lines represent PDOS of C-2p, C-2s, K-4s, Bi-6s, and Bi-5p orbitals, respectively. }
\label{Figure4}
\end{figure}

To identify the complicated crystal structure for the 3.5~K SC phase, we first employed the Universal Structure Predictor:
Evolutionary Xtallography (USPEX) based on the genetic algorithm~\cite{USPEX2006} to search for global stable or metastable structures
in the phase diagram of K$_y$Bi with $y$=1-4, which correspond to the possible arrangement of potassium and bismuth atoms in the
doped materials. In the search process, the plane-wave pseudopotential method as implemented in the Vienna $ab$ initio simulation package
(VASP) program~\cite{Kresse1993,Kresse1996} was adopted to relax the atomic positions. The generalized gradient approximation (GGA)
with Perdew-Burke-Ernzerhof (PBE) formula~\cite{Perdew1996} for the exchange-correlation potentials and the projector-augmented wave
method (PAW)~\cite{Blochl1994} for ionic potential were used to model the electron-electron and electron-ion interactions.
The searched results indicate that one cubic structure of K$_4$Bi could reflect the main character of measured XRD pattern. Then we replace
bismuth atom with triphenylbismuth in this structure and perform a full relaxation of atomic positions. The optimized crystal has
three molecules of C$_{18}$H$_{15}$Bi and twelve K atoms in a unit cell of dimensions a=b=c=9.473~$\AA$, and $\alpha=\beta=\gamma=90^{o}$,
as shown in Fig.~\ref{Figure3}b. Potassium atoms represented by blue balls in Fig.~\ref{Figure3}b are intercalated in the interstitial
space of bismuth and phenyl rings, with deep blue one close to a certain phenyl ring and light blue one occupying the center of green
dashed line connecting two bismuth atoms. The powder XRD pattern based on the optimized crystal, shown at the bottom of Fig.~\ref{Figure2},
is in good agreement with the one marked by $*$. This indicates that the doped materials crystallize into the optimized cubic structure
with high probability. The missing of weak peaks in K$_x$TPB ($x$=1-3) compared with the theoretical modeling is mostly possible due to
poor crystallization, manifested by broad XRD peaks in Fig.~\ref{Figure2}.

The orbital-resolved partial density of states (PDOS) for the optimized crystal are presented in Fig.~\ref{Figure4}. The existence of
finite PDOS at the Fermi energy indicates that the potassium-doped system actually lies in the metallic state, providing a support for
the observation of superconductivity at 3.5~K. Among the five orbital shown in Fig.~\ref{Figure4}, the C-2p orbital makes a dominant
contribution to PDOS in the vicinity of the Fermi energy, while the K-4s orbital has little contribution. This result reflects the fact
that the electron is transferred from K-4s to C-2p, which not only leads to the metallic behavior but also strongly affects the
vibrations of phenyl rings.

\begin{figure}
\includegraphics[width=0.48\textwidth]{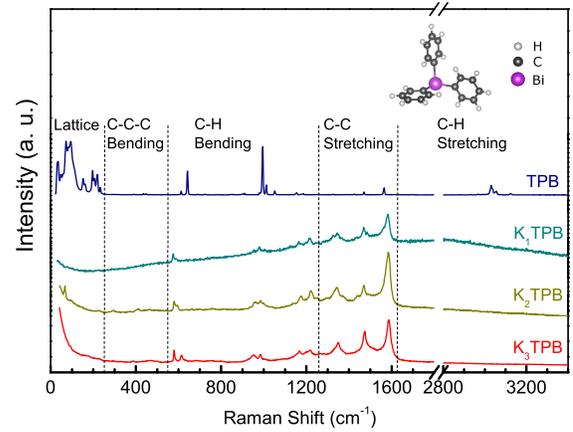}
\caption{(Color online) Raman scattering spectra of pristine and potassium-doped triphenylbismuth collected at room temperature. Upper right
presents the molecular structure of triphenylbismuth. Five regions of Raman active modes, divided by the vertical dashed lines, are shown above
the spectra of pristine material.}
\label{Figure5}
\end{figure}

\textit{Raman spectra of pristine and potassium-doped triphenylbismuth--}
The SC phase was further characterized by phase-sensitive Raman spectra, which were collected on an in-house system with Charge
Coupled Device and Spectrometer from Princeton Instruments. Five regions of Raman active modes from the low to high frequencies
correspond to the lattice and Bi-phenyl, C-C-C bending, C-H bending, C-C stretching, and C-H stretching modes~\cite{Ludwig1995}.
We observed all these modes in pristine triphenylbismuth. Upon doping potassium into triphenylbismuth, all lattice modes are dramatically
suppressed and the modes in the C-H stretching region become invisible (see Figure~\ref{Figure5}).

Significant differences of the spectra between the pristine and doped samples are in the C-H bending and C-C stretching regions. Upon doping
potassium, the 644 and 993~cm$^{-1}$ C-H bending modes in the pristine material shift down with a dramatic decrease in the intensity.
By contrast, the C-H bending modes at 1154 and 1182~cm$^{-1}$ shift up with an increase in the intensity. It is obvious that the mode intensity
in the whole C-C stretching region gets a strong enhancement in the doped samples. An upshift of Raman spectra is also observed for the two
peaks at 1322 and 1564~cm$^{-1}$ in the pristine material, while the peak at 1469~cm$^{-1}$ does not shift its position with the potassium doping.

The observation of both red and blue shifts of Raman spectra in potassium-doped triphenylbismuth is quite different from the situation in
potassium-doped phenanthrene~\cite{XFWang2011} and picene~\cite{Kambe2012}, where only red shifts were observed. Such red shifts were
attributed to the softening of Raman modes by the transferred electrons from potassium to phenanthrene and picene molecules. This mechanism
should also work for triphenylbismuth, making the Raman modes tend to shift down. On the other hand, when phenyl is connected to an X
(either metal~\cite{Shobatake1969} or halogen~\cite{Whiffen1956}) atom, both the C-H and C-C modes are affected by the C-X bond. In the
halogeno-benzenes~\cite{Whiffen1956}, the frequencies of the C-C stretching modes increase from I-benzene to
F-benzene, suggesting that an increase of benzene polarization with increasing the electronegativity of halogen atom makes the Raman modes
tend to shift up. Based on the above analyses, the Raman shifts in our samples could be understood as the competing results between transferred
electrons and enhanced polarization of phenyls, which is clearly manifested by the asymmetric Raman line shape and the increase of Raman
intensity in the C-C stretching region.

\textit{Conclusion--}
The present results provide unambiguous evidence for highly reproducible superconductivity in potassium-doped triphenylbismuth.
The existence of 3.5~K and 7.2~K SC transitions indicates that two stable crystal structures can be formed upon doping potassium into
triphenylbismuth. Similar SC transitions around 7.2~K were also observed in potassium-doped picene~\cite{Mitsuhashi2010},
dibenzopentacene~\cite{Xue2012}, and $p$-terphenyl~\cite{RSWang2017}. The insensitivity of T$_c$ to the details of molecules suggests
that the SC phase around 7.2~K rests mainly on the common structural unit - phenyl ring. On the other hand, the findings of SC phases
with much higher T$_c$'s (18~K in picene~\cite{Mitsuhashi2010}, 33~K in dibenzopentacene~\cite{Xue2012},
and above 120~K in $p$-terphenyl~\cite{RSWang2017}) indicate that the arrangement of phenyl rings plays a crucial role in boosting
superconductivity.

The unique feature of triphenylbismuth-like organometallic molecules is the presence of metal atom connecting phenyl rings. Since the structure
of such systems depends sensitively on the metal atom, finding materials with different T$_{c}$'s is expected, which provides a bottom-up
understanding of the SC mechanism in the phenyl-ring-based organic superconductors.

R.S.W. and J.C. contribute equally to this work. We thank Hai-Qing Lin for strong support and valuable discussion. This work was supported
by the National Natural Science Foundation of China under Grants Nos.~11574076, 11674087, and 91221103.


\end{document}